\documentclass[a4paper,twocolumn]{article}

\usepackage[utf8]{inputenc}
\usepackage[T1]{fontenc}
\usepackage{amsmath, bm, commath, mathtools, amssymb}
\usepackage{hyperref}
\usepackage[capitalise]{cleveref}
\usepackage{siunitx, booktabs, pbox}
\usepackage[table]{xcolor}

\usepackage{authblk}

\usepackage{natbib}
\setcitestyle{numbers}

\hypersetup{colorlinks = true}

\newcommand{\Rossby}{\mbox{\textit{Ro}}}

\newcommand{\ReynoldsM}{\mbox{\textit{Rm}}}
\newcommand{\Ekman}{\mbox{\textit{Ek}}}
\newcommand{\Prandtlm}{\mbox{\textit{Pm}}}

\newcommand{\myvec}[1]{\bm{#1}}
\newcommand{\myck}{\myvec{K}}
\newcommand{\conj}[1]{{#1}^{*}}

\newcommand{\myk}{k}
\newcommand{\avg}[1]{\overline{#1}}

\begin{document}
	\title{Magnetic structure, dipole reversals, and 1/f noise in resistive MHD spherical dynamos}
	\author{M. Fontana\footnote{mfontana@df.uba.ar}}
	\author{P. D. Mininni}
	\author{P. Dmitruk}
	\affil{Departamento de Física, Facultad de Ciencias Exactas y Naturales, Universidad de Buenos Aires, and IFIBA, CONICET, Ciudad Universitaria, 1428 Buenos Aires, Argentina.}

\twocolumn[
\begin{@twocolumnfalse}
	\maketitle
	\begin{abstract}
		A parametric study of the magnetic dipole behavior in resistive incompressible MHD inside a rotating sphere is performed, using direct numerical simulations and considering Reynolds and Ekman numbers as controlling parameters. The tendency is to obtain geodynamo-like magnetic dipole reversal regimes for sufficiently small Ekman and large Reynolds numbers. The typical dipole latitude obtained in the reversal regime is around 40 degrees (with respect to the rotation axis of the sphere). A statistical analysis of waiting times between dipole reversals is also performed, obtaining a non-Poissonian distribution of waiting times, indicating long-term memory effects. We also report the presence of a $1/f$ frequency power spectrum in the magnetic dipole time-series, which also shows a tendency to grow toward lower frequencies as the Ekman number is decreased.
		\vspace{1.5cm}
	\end{abstract}
	\end{@twocolumnfalse}
]

\section{Introduction}
Several naturally occurring magnetic fields, such as those found in stars and planets, are thought to be sustained by magnetohydrodynamic dynamos, i.e., by the induction resulting from the flow of highly conducting materials in their interiors. This hypothesis is supported by numerous theoretical, experimental, and numerical works \cite{Glatzmaier1995,Monchaux2009,Christensen2010,Driscoll2016}. However, due to the chaotic (and often turbulent) nature of these type of systems, and the extreme regimes in which they operate, a completely accurate experimental or numerical reproduction has been, so far, unattainable.

A distinct characteristic of some of these dynamos is that their dipole moment rapidly reverses its direction, a process known as polarity inversion or reversal. In the case of the Sun, periodic polarity inversions take place every 11 years \cite{Vecchio2012}. On the other hand, geomagnetic reversals occur in a much longer and broader set of timescales, of the order of $\sim 10^5-10^7$ years \cite{Laj2007,Amit2010}, and their observation requires the measurement of the remanent magnetization of rocks found on Earth's crust. Recent dynamo experiments \cite{Nornberg2006,Verhille2010,Berhanu2010} and numerical simulations \cite{Christensen1999,Rotvig2002,Morin2009,Reuter2009,Mininni2014} have shown that both homogeneous and inhomogeneous dynamos constrained in different geometries display a rich set of bifurcations, including non-dynamo regimes, intermittent, periodic, quasi-periodic, and aperiodic dynamos. In these studies, the controlling parameters that determined the operating regime were the Ekman, Rossby, Reynolds, and magnetic Reynolds numbers. The particular importance of the Ekman and magnetic Reynolds numbers has also been pointed out in the specific context of the geodynamo \cite{Christensen1999,Rotvig2002,Morin2009}, given the extreme values these parameters can take in the Earth's core, and their role in numerical simulations that can capture field reversals in the so-called strong field regime.

The aperiodic reversals observed in the geodynamo are known to deviate from a purely Poisson process and showcase long-term memory \cite{Carbone2006}. Such statistics have also been found in ideal hydrodynamic (HD) and magnetohydrodynamic (MHD) simulations, and linked to the presence of $1/f$ noise in the temporal power spectrum of the system energy \cite{Dmitruk2007,delaTorre2007,Dmitruk2011}. In a previous publication, we showed the presence of $1/f$ noise in the power spectrum of the dipole moment for the case of ideal MHD simulations in spherical vessels \cite{Dmitruk2014}, a feature also found in actual paleomagnetic data \cite{Constable2005,Ziegler2011}. More recently, we also showed that $1/f$ noise and long term memory also develops in a dynamo experiment in a cylindrical vessel, as well as in numerical simulations of MHD dynamos using a flow similar to that used in the experiments \cite{Mininni2014}.

In the present work we consider the effect of the Ekman and Reynolds numbers in the development of dynamo solutions displaying field reversals, long term memory, and $1/f$ dynamics. We solve the incompressible MHD equations inside a full sphere using direct numerical simulations. Only dissipative and Coriolis effects are considered. Hence, our model cannot be compared with the great efforts behind geodynamo simulations \cite{Jones2011,Matsui2016,Schaeffer2017}, which take into account more complex physics, such as compressibility effects or chemical reactions between constituent phases. We argue, however, that this simplified system suffices to reproduce some statistical features of the geomagnetic dipole moment, indicating also that different solutions exist in parameter space depending on the Ekman and magnetic Reynolds numbers. Finally, by employing a fully spectral scheme in space, we manage to simulate up to 2,000 eddy turnover times while maintaining low numerical dispersion. This, in turn, allow us to provide further evidence about the link between long-term memory, $1/f$ noise and geomagnetic reversals.

\section{Simulations setup}
\subsection{Model equations and numerical scheme}
For this work, we analyze a domain $\Gamma$ consisting of a unit sphere filled with an incompressible conducting fluid where a mechanical forcing field $\myvec f$ (assumed to be solenoidal) is acting. The density of the fluid is taken to be uniform and equal to unity and the boundary of the sphere is considered to be rotating with angular speed $\Omega$. For this scenario, the dynamics are more easily represented in a non-inertial reference frame which rotates with the domain's border, and which can be described by the usual magnetohydrodynamic equations:
\begin{gather}
\label{eq:navier-stokes}
\begin{split}	
\partial_t \myvec v = - \myvec \nabla \mathcal P + \myvec v \times \myvec \omega + \myvec j \times \myvec b - 2\Omega {\myvec e}_z \times \myvec v +\\
+ \nu \nabla^2 \myvec v + \myvec f,
\end{split}\\
\label{eq:induction}
\partial_t \myvec b = \myvec \nabla \times \left ( \myvec v \times \myvec b \right) + \eta \nabla^2 \myvec b,\\
\label{eq:incompressible}
\myvec \nabla \cdot \myvec v = 0,\\
\label{eq:magnetic-gauss}
\myvec \nabla \cdot \myvec b = 0,
\end{gather}
where $\myvec v$ is the velocity field, $\myvec \omega = \myvec \nabla \times \myvec v$ the vorticity, $\myvec b$ the magnetic field, $\myvec j=\myvec \nabla \times \myvec b$ the current density, $\mathcal P$ the total pressure, $\nu$ the kinematic viscosity, $\eta$ the magnetic diffusivity, and ${\myvec e}_z$ is a unit vector in the direction of the angular velocity. For simplicity, all quantities are expressed in an alfvenic unit system. 

To solve \cref{eq:incompressible,eq:induction,eq:magnetic-gauss,eq:navier-stokes}, suitable boundary conditions must be supplied. We consider the homogeneous Neumann boundary conditions case ($\myvec b \cdot \myvec {\hat n} = \myvec j \cdot \myvec {\hat n} = \myvec v \cdot \myvec {\hat n} = \myvec \omega \cdot \myvec {\hat n} = 0$, where $\myvec {\hat n}$ represents a unit vector normal to the sphere surface). Physically, this represents an impenetrable, perfect conductor at the boundary, which actually entails $\myvec{b} = \myvec{0}$ on $\partial \Gamma$. Moreover, this conductor is coated on the inside with a thin layer of dielectric material. The vanishing of the normal component of $\myvec \omega$ is implied by, but does not imply, no-slip boundary conditions.

It is a well-known fact that the aforementioned system of equations can be solved analytically only for some rather simple scenarios, mainly due to the presence of non-linear terms. Hence, for a more general exploration of parameter space, we numerically integrate \cref{eq:incompressible,eq:induction,eq:magnetic-gauss,eq:navier-stokes} for hundreds of turnover times. To accomplish this, the fields are decomposed using Chandrasekhar-Kendall (CK) eigenfunctions of the curl as the spectral basis, as reported in \cite{Mininni2006,Mininni2007}. By definition, the elements of the basis $\myck_i$ must satisfy 
\begin{equation}
\myvec \nabla \times \myck_i = \myk_i \myck_i.
\label{eq:ck-definition}
\end{equation}
Although the general solvability of \cref{eq:ck-definition} is an open mathematical problem, it has been demonstrated that its solutions for the linear case (uniform $\myk_i$) provide a basis for divergenceless fields with homogeneous Neumann boundary conditions \cite{Yoshida1990}.

Eigenfunctions $\myck_i$ can be found by transforming \cref{eq:ck-definition} into a Helmholtz equation, whose solutions depend on three indexes $q, l, m$, and can be shown to be obtained as $\myck_{qlm} = \myk_{ql} \myvec \nabla \times \left ( \myvec \nabla \times \psi_{qlm} \myvec {\hat r} \right ) + \myvec \nabla \times \psi_{qlm} \myvec {\hat r}$. Here, $\psi$ is a solution to the scalar Helmholtz equation which, for spherical coordinates and in a full sphere, is given by
\begin{equation}
\psi_{qlm} = C_{ql} j_l (|\myk_{ql}| r) Y_l^m (\theta, \phi).
\end{equation}
In this equation $j_l$ represents the $l$-th order spherical Bessel function and $Y_l^m$ is the usual spherical harmonic of degree $l$ and order $m$, as a function of polar angle $\theta$ and azimuthal angle $\phi$. $C_{ql}$ is a normalization constant, appropriately chosen so that the eigenfunctions are orthonormal with respect to the standard inner product. The index $q$ can take any integer value except for zero; $l$ and $m$ follow the usual rules for spherical harmonic indexing ($l>1$, $-l<m<l$).

An equation for the double infinite series of eigenvalues $\myk_{ql}$ is easily derived using the boundary conditions. These eigenvalues satisfy $\myk_{-ql} = \myk_{ql}$, and can be interpreted physically as an analog of the wavenumber used in Fourier decompositions. A consequence of the previous relation is that eigenfunctions $\myck_{qlm}$ and $\myck_{-qlm}$ differ only on the sign of their helicity. 

The fields can then be expressed as
\begin{align}
\myvec v (\myvec r, t) &= \sum_{q=-\infty}^{\infty} \sum_{l=1}^{\infty} \sum_{m=-l}^{m=l} \xi_{qlm}^v (t) \myck_{qlm} (\myvec r),
\label{eq:v-ck} \\
\myvec b (\myvec r, t) &= \sum_{q=-\infty}^{\infty} \sum_{l=1}^{\infty} \sum_{m=-l}^{m=l} \xi_{qlm}^b (t) \myck_{qlm} (\myvec r),
\label{eq:b-ck}
\end{align}
with an analogous expression for $\myvec f$. It is to be noted that, as all the fields are real, the relation $\xi_{ql-m} = (-1)^{m}\conj{\xi}_{qlm}$ must hold for all the coefficients, where the asterisk denotes the complex conjugation operation. By substituting \cref{eq:v-ck,eq:b-ck} in \cref{eq:navier-stokes,eq:induction}, using \cref{eq:ck-definition} and the orthogonality of the eigenfunctions, it is possible to obtain ODEs for the time evolution of the expansion coefficients $\xi$ for each field: 
\begin{align}
\dod{\xi_n^v}{t} =& \sum_{i,j} \myk_j I_{ij}^n \left( \xi_i^v \xi_j^v - \xi_i^b \xi_j^b \right) + 2 \Omega \sum_i \xi_n^v \myvec {e}_z \cdot \myvec O_i^n - \nonumber \\
&-\nu \myk^2 \xi_n^v + \xi_n^f, 
\label{eq:ck-navier-stokes}\\
\dod{\xi_n^b}{t} =& \sum_{i,j} \myk_n I_{ij}^n \xi_i^v \xi_j^b - \eta \myk^2 \xi_n^b,
\label{eq:ck-induction}
\end{align}
where each of $i, j$ and $n$ represents a $(q, l, m)$ triplet. $I_{ij}^n$ and $\myvec O_i^n$ are coupling arrays defined as
\begin{align}
I_{ij}^n &= \int_\Gamma \conj {\myck_n} \cdot \left( \myck_i \times \myck_j \right) \dif \Gamma, \\
\myvec O_i^n &= \int_\Gamma \conj{\myck_n} \times \myck_i \dif \Gamma.
\end{align}

For numerically solving \cref{eq:ck-navier-stokes,eq:ck-induction}, first a spectral resolution $q_\text{max}, l_\text{max}$ must be chosen. Then, the computation of the normalization constants $C_{ql}$, the eigenvalues $\myk_i$, and the coupling arrays $I_{ij}^n, \myvec O_i^n$ must be performed. Finally, the coefficients $\xi^v$ and $\xi^b$ are evolved in time using, in our case, a 4th-order Runge-Kutta scheme. The code is parallelized using MPI. It is to be noted that, although the proposed decomposition and numerical scheme has been proved highly accurate \cite{Mininni2006} and scales well with number of processors, the absence of a fast spherical Bessel transform algorithm considerably limits the regions of parameter space that can be explored. However, the method has some advantages that are worth pointing out in the context of this study. First, it allows computations in full spheres without a numerical singularity at the origin. Second, as the method is spectral and thus conserves the quadratic invariants of the system, it allows for long time integrations with small accumulation of errors \cite{Mininni2006}.

\begin{table*}[t!]
	\sisetup{table-format=3.0, table-column-width=0.15\textwidth}
	\rowcolors{2}{gray!20}{white}
	\begin{tabular}{c *3{S[table-format=-1.2e2, retain-zero-exponent=true]} *2{S[table-format=-1.2]}}
		\toprule
		\toprule
		ID & $\ReynoldsM$ & $\Ekman$ & $\Rossby$ & {Avg. Dipolarity} & {Avg. Abs. Latitude}\\
		\midrule
		\midrule
		ND01 & 6.11e+01 & 1.95e-05 & 1.19e-03 & {--} & {--} \\
		ND02 & 6.21e+01 & 3.91e-05 & 2.42e-03 & {--} & {--} \\
		ND03 & 7.14e+01 & 1.56e-04 & 1.12e-02 & {--} & {--} \\
		STAT01 & 9.81e+01 & 9.38e-05 & 9.20e-03 & 37.61 & 72.07\\
		STAT02 & 1.28e+02 & 6.25e-05 & 8.02e-03 & 37.29 & 75.46\\
		STAT03 & 1.64e+02 & 1.20e-05 & 1.97e-03 & 0.13 & 48.48\\
		REV01 & 2.12e+02 & 2.50e-04 & 5.31e-02 & 29.44 & 20.44\\
		REV02 & 2.25e+02 & 4.69e-05 & 1.05e-02 & 14.48 & 33.68\\
		REV03 & 2.47e+02 & 1.50e-03 & 3.71e-01 & 26.60 & 26.40\\
		REV04 & 2.69e+02 & 1.88e-04 & 5.05e-02 & 29.31 & 23.33\\
		REV05 & 2.91e+02 & 3.13e-05 & 9.10e-03 & 15.39 & 35.59\\
		REV06 & 3.03e+02 & 6.25e-05 & 1.89e-02 & 22.80 & 27.04\\
		REV07 & 3.90e+02 & 1.00e-03 & 3.90e-01 & 21.64 & 28.33\\
		REV08 & 4.27e+02 & 4.69e-05 & 2.00e-02 & 18.51 & 25.61\\
		REV09 & 4.32e+02 & 1.56e-05 & 6.75e-03 & 15.00 & 42.09\\
		REV10 & 4.91e+02 & 6.64e-06 & 3.26e-03 & 12.63 & 41.04\\
		REV11 & 5.01e+02 & 3.13e-04 & 1.56e-01 & 24.13 & 28.39\\
		REV12 & 5.32e+02 & 5.83e-05 & 3.10e-02 & 18.80 & 28.79\\
		REV13 & 5.48e+02 & 7.50e-04 & 4.11e-01 & 19.70 & 29.14\\
		REV14 & 5.99e+02 & 7.81e-06 & 4.68e-03 & 9.76 & 35.57\\
		REV15 & 6.36e+02 & 1.50e-03 & 9.55e-01 & 13.57 & 41.45\\
		REV16 & 6.78e+02 & 2.50e-04 & 1.69e-01 & 19.04 & 32.41\\
		REV17 & 7.36e+02 & 3.13e-05 & 2.30e-02 & 11.74 & 38.81\\
		REV18 & 1.23e+03 & 6.25e-05 & 7.69e-02 & 14.82 & 29.69\\
		SS01 & 6.66e+02 & 6.00e-03 & 4.00e+00 & 7.19 & 29.87\\
		SS02 & 1.06e+03 & 1.25e-03 & 1.32e+00 & 6.66 & 34.50\\
		SS03 & 1.07e+03 & 6.25e-03 & 6.70e+00 & 6.01 & 32.95\\
		SS04 & 1.07e+03 & 4.00e-03 & 4.29e+00 & 6.10 & 31.93\\
		SS05 & 2.22e+03 & 2.50e-04 & 5.55e-01 & 5.76 & 33.86\\
		SS06 & 2.51e+03 & 1.56e-03 & 3.93e+00 & 4.96 & 32.94\\
		\bottomrule
		\bottomrule
	\end{tabular}
	\caption{Summary of runs with their respective names (ID) and associated adimensional parameters: Magnetic Reynolds number ($Rm$), Ekman ($Ek$) and Rossby ($Ro$) numbers (all calculated at the sphere radius), average values of dipolarity and of absolute latitude of the dipole moment (i.e., its orientation with respect to the angular velocity). The alphabetic part of the ID follows the classification introduced in sec. III.A.}
	\label{tab:parameters}
\end{table*}
\begin{figure*}
	\centering
	\includegraphics[width=.80\textwidth, height=.26\textwidth]{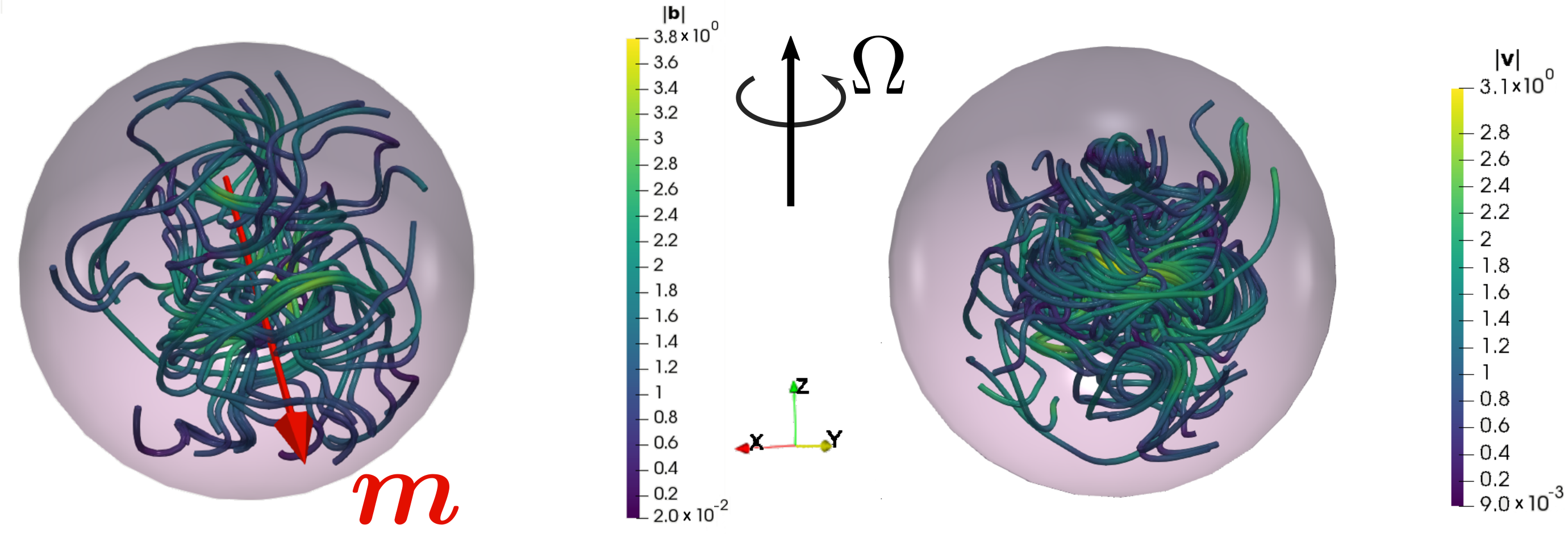}
	\caption{Magnetic field lines (left) and streamlines (right) for simulation REV13 at $t=1980$. Both sets of lines are colored according to the local intensity of the respective field and the magnetic dipole moment's direction is marked with a red arrow. The angular speed $\bm \Omega$ and the orientation of unit vectors $\bm{\hat x}$, $\bm{\hat y}$, $\bm{\hat z}$ are also shown}.
	\label{fig:fields}
\end{figure*}

\subsection{Parameters}
To probe parameter space we employ over thirty direct numerical simulations, carried over using double precision arithmetic and with a resolution $q_\text{max}=l_\text{max}=7$ (i.e., with 977 expansion coefficients). This corresponds to an equivalent maximum "wavenumber" of $k_\text{max}\approx32$. Although somewhat low, this choice allows us to simulate up to 400 large-scale eddy turnover times in each simulation (and up to $2,000$ for a subset of runs discussed below), with a temporal resolution $\Delta t=1\times 10^{-3}$. The forcing $\myvec f$ and the initial conditions are the same in all the cases. In spectral domain, the former is chosen to be represented as
\begin{equation}
\xi^f_{qlm} = 
\left\{
\begin{aligned}
&3.5\quad&&\text{if} \quad q=l=3, \ m>0,\\
&(-1)^m \times 3.5 \quad&& \text{if} \quad q=l=3, \ m<0,\\
&0 \quad&&\text{if} \quad q\ne 3, \ l\ne 3.
\end{aligned}
\right.
\end{equation}

This constitutes a steady forcing field with positive net helicity (only positive values of $q$ are excited) and which injects kinetic energy at an intermediate scale, properties known to favor the generation of large-scale magnetic fields \cite{Parker1970,Pouquet1976}. The $3.5$ factor is chosen ad-hoc in order to approximately normalize the average kinetic energy in the turbulent steady state to a value of order unity. We have verified that other choices for the forcing coefficients (with the same overall properties) yield similar results. Initial conditions for the magnetic and velocity fields consist on exciting only the $q=l=1$ modes for the former, and all $q=l=1$ and $q=l=3$ modes for the latter.

\begin{figure*}[t]
	\centering
	\includegraphics[width=.98\textwidth, keepaspectratio=true]{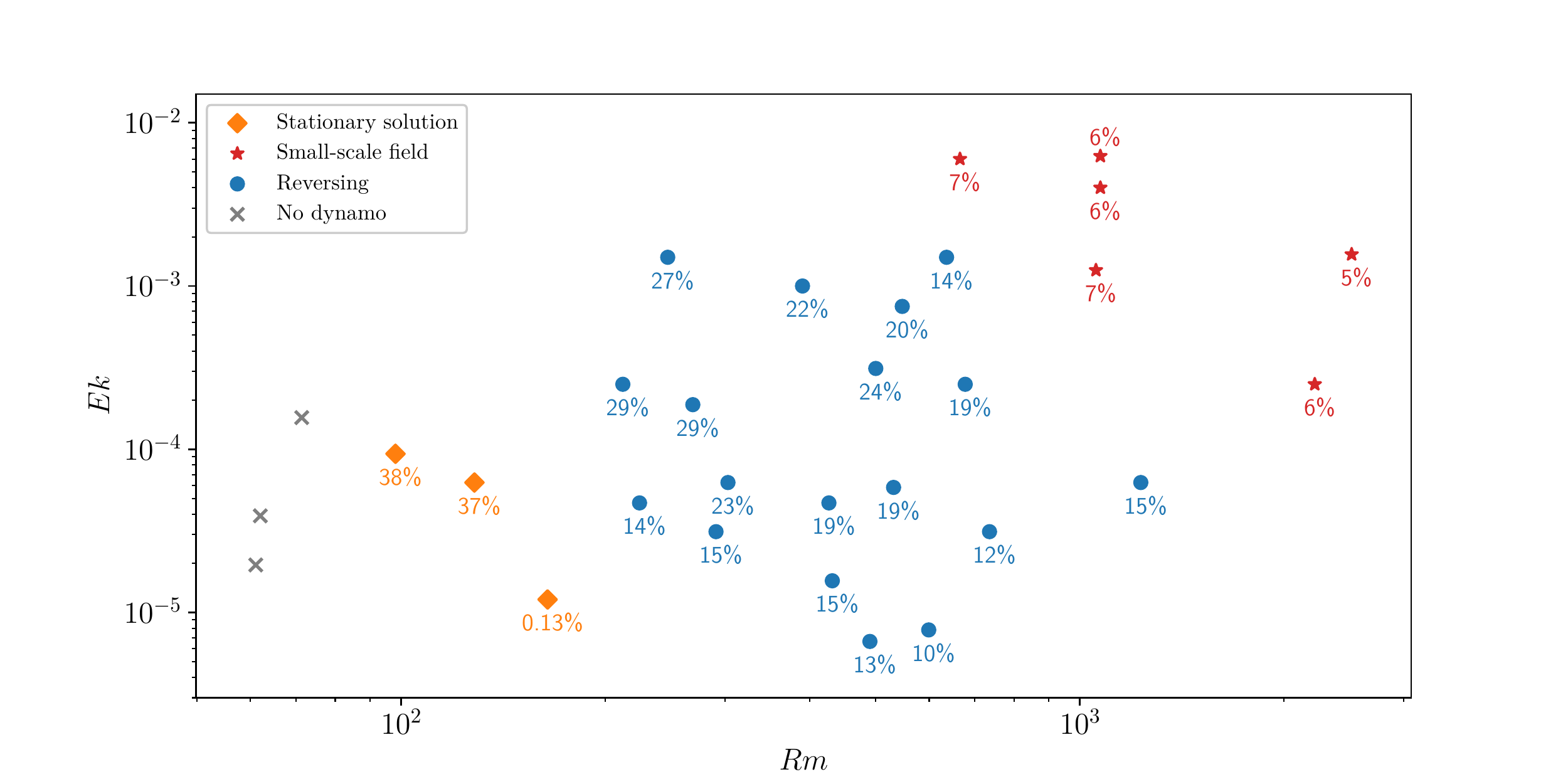}
	\caption{Numerical simulations performed and their classification depending on the type of dynamo solution, as a function of Ekman ($\Ekman$) and magnetic Reynolds ($\ReynoldsM$) numbers. Below the markers, corresponding time-averaged dipolarity percentage are specified.}
	\label{fig:overview}
\end{figure*}

The input parameters that vary across simulations are the angular speed $\Omega$, the kinematic viscosity $\nu$, and the magnetic diffusivity $\eta$. In all the cases a unit magnetic Prandtl number $\Prandtlm$ is imposed (i.e. $\nu=\eta$). Although distant from the actual value of $\Prandtlm$ for geodynamo's, this choice is necessary to keep simulations attainable with present computational resources (also, note that given the relatively small resolution used, if $\nu$ and $\eta$ are interpreted as turbulent transport coefficients, then the effective turbulent magnetic Prandtl number can be expected to be of order unity in a turbulent dynamo regime \cite{Ponty2005}). The results are then analyzed as a function of each simulation's Ekman number $\Ekman$ and Magnetic Reynolds number $\ReynoldsM$ (equal to the kinetic Reynolds number as $\nu=\eta$). These can be calculated as 
\begin{equation}
\ReynoldsM = \frac{\avg{v} R}{\eta}, \qquad \qquad \Ekman= \frac{\nu}{R^2 \Omega},
\end{equation}
where $R$ is the radius of the sphere (equal to unity in dimensionless units) and $\avg{v}$ is the time-averaged intensity of the velocity field.

One way to analyze the magnetic field's degree of symmetry is to calculate the magnetic energy of each spherical harmonic degree $E^b_l$,
\begin{equation}
E^b_l = \frac{1}{2} \sum_{q,m} |\xi_{qlm}^b|^2,
\end{equation}
with $E^b_1$ corresponding to the magnetic energy contribution from the dipolar field (referred as dipolar energy for simplicity), $E^b_2$ the quadrupolar energy, and so on. The fraction of the magnetic energy in the dipolar field $E^b_1$ to the total magnetic energy $E^b$, in percentage, will be referred to in the following as the "average dipolarity." Moreover, the magnetic dipole moment vector $\myvec{m}$,
\begin{equation}
\myvec m = \int_\Omega \myvec r \times \myvec j \dif \Omega,
\end{equation}
can also be obtained from the $\xi^b_{q1m}$ coefficients and, from it, the latitude of the dipole moment $\alpha$ can be computed as
\begin{equation}
\alpha = \arctan \left( m_z \Big / \sqrt{m_x^2 + m_y^2} \right ).
\end{equation}

\section{Results}
We report here thirty runs, a summary of which is presented in table \ref{tab:parameters}. The name (ID) of each run, and their adimensional parameters (the magnetic Reynolds, Ekman and Rossby numbers, all calculated at the largest scale of the system) are included. The average values of dipolarity and absolute latitude of dipole moment, as discussed in the previous section, are also reported.

\subsection{Magnetic field structure}
A qualitative view of the magnetic field for a specific run (REV13) is shown in \cref{fig:fields}, where we also depict the corresponding velocity field streamlines and the dipole orientation. This is to illustrate the typical complex structure of the fields that can be found in most of the runs. We also report an overview of the magnetic field structure as a function of \emph{Magnetic Reynolds} ($\ReynoldsM$) and \emph{Ekman} ($\Ekman$) numbers. A graphical summary of these findings is shown in \cref{fig:overview}, where the following classification is employed:
\begin{itemize}
	\item \textbf{\textit{No-dynamo region} (ND runs)}: As in previous studies in different geometries and with varying dynamo configurations \cite{Gubbins1973,Leorat1981,Christensen2010,Monchaux2009}, we find a \emph{Critical Magnetic Reynolds} which imposes a constraint on dynamo action ($\ReynoldsM_\text{crit} \approx 80$ in our simulations) . For values of $\ReynoldsM$ lower than $\ReynoldsM_\text{crit}$ the system converges to the extinction of $\myvec{b}$ in just a few turnover times.
	
	\item \textbf{\textit{Stationary solutions} (STAT runs)}: In the case of simulations with $\ReynoldsM$ slightly above $\ReynoldsM_\text{crit}$, the solutions obtained seem to asymptotically converge towards a stationary state in all the cases. 

	\item \textbf{\textit{Reversals} (REV runs)}: For $\ReynoldsM \approx 2 \times 10^2$ and $Ek \lesssim 2\times10^{-3}$, the system starts exhibiting time variability and several polarity inversions of the magnetic dipole  moment are found.
	
	\item \textbf{\textit{Small scale field} (SS runs)}: Further augmenting $\ReynoldsM$, while also increasing $\Ekman$, results in a transition towards a regime where the growth of large-scale magnetic fields becomes negligible. In this regime $\myvec b$ is mostly defined by its energy at scales smaller than the energy containing scale of the flow.
\end{itemize}

The classification proposed above broadly depicts the dynamics found in each region of parameter space. Even though the geometry of the vessel and of the flow are different, it is interesting that qualitatively similar behaviors were reported in explorations of parameter space in laboratory dynamo experiments \cite{Berhanu2010}. In a previous work with the same geometry \cite{Mininni2007} quasi-periodic solutions were also found in a region of parameter space not explored in this study, with low Rm and small Ek (it is interesting that these solutions are also observed in the laboratory dynamo experiments). To better quantify the generated magnetic field's degree of symmetry in our solutions, the average dipolarity of each simulation was computed (also shown in \cref{fig:overview}). As a general rule, higher values of $\ReynoldsM$ seem to generate less dipolar dynamos. The role of the Coriolis force, on the other hand, seems more complex, with higher values of $\Ekman$ leading to larger dipolarities right until the proximity of the small-scale field transition, where this trend is reversed. 

\begin{figure}[t]
	\centering
	\includegraphics[width=.49\textwidth,keepaspectratio=true]{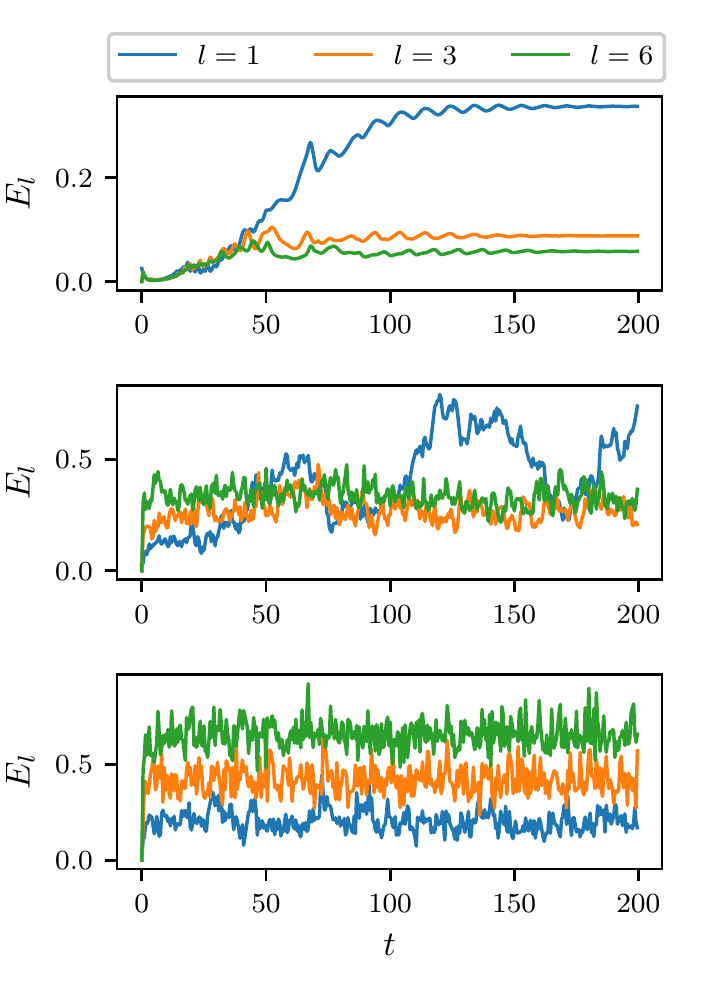}
	\caption{Magnetic energy per spherical harmonic degree $l$, integrated over the entire sphere, for times between 0 and 200. For clarity, only three modes are displayed, namely $l=1$ (dipolar energy), $l=3$ and $l=6$. The plots correspond to simulations STAT01 (top), REV12 (center) and SS03 (bottom).}
	\label{fig:mspectrums}
\end{figure}

For the simulations with stationary solutions, the effects of $\Ekman$ on the magnetic field structure appear to be more abrupt. In that case, the system shows a high sensitivity to the operating Ekman number, as a sharp decrease of more than two orders of magnitude in the dipolarity is found while diminishing $\Ekman$ by a factor of ten (see also table \ref{tab:parameters}). 

Although in all the cases simulations tagged as reversing display larger average dipolarities, the criterion that separates them from the small-scale field simulations relies upon the time variability of the more energetic magnetic modes. To better illustrate this, the magnetic energy per scale is plotted as a function of time for three distinct simulations, one per each dynamo region in parameter space, in \cref{fig:mspectrums}. As a measure of scale, the spherical harmonic degree $l$ was chosen. In these graphs, the difference between simulations classified as reversing and small-scale field becomes clear: the latter display a marginal magnetic dipolar energy at every time, whereas the former exhibit important periods of magnetic dipolar energy being the most energetic. Also to be noted in \cref{fig:mspectrums} is the convergence towards a steady value of the magnetic energy in simulations with $\ReynoldsM$ slightly over $\ReynoldsM_\text{crit}$.

\begin{figure}[t]
	\centering
	\includegraphics[width=.45\textwidth,keepaspectratio=true]{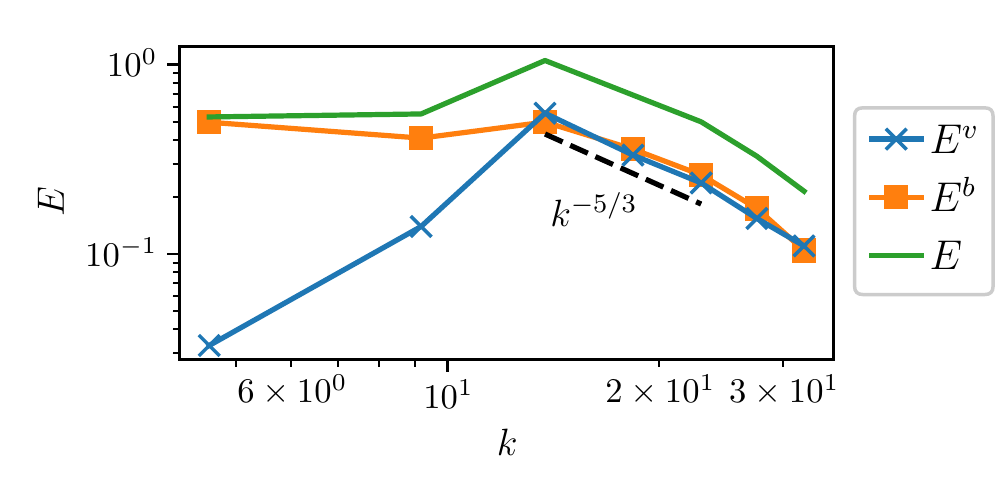}
	\caption{Kinetic (blue), magnetic (orange) and total (green) energy spectra for run REV11, averaged between $t=300$ and $t=400$. A $k^{-5/3}$ power law is also shown for comparison.}
	\label{fig:kspectrum}
\end{figure}

Another important property that characterizes dynamo regimes is the ratio of magnetic versus kinetic energy. The spectra of both of this quantities are shown in \cref{fig:kspectrum}, as a function of the wavenumber $k$, for a run with typical $\ReynoldsM$ and $\Ekman$ values. As expected, equipartition of energy is observed at  intermediate and small scales (i.e., at intermediate and small wavenumbers). However, at the largest scales, an inverse transfer of magnetic energy is observed, and the system's energy is predominantly magnetic. Also, albeit with limited spatial resolution, fluctuations in the velocity and magnetic field can be seen below the forcing scale, and we show as a reference a Kolmogorov $k^{-5/3}$ power law. It is to this state that we loosely referred to previously in the text as a \textit{turbulent steady state}, i.e., as a state in which fluctuations at multiple scales (not excited directly by our forcing) are present.

\begin{figure}[t]
	\centering
	\includegraphics[width=.45\textwidth,keepaspectratio=true]{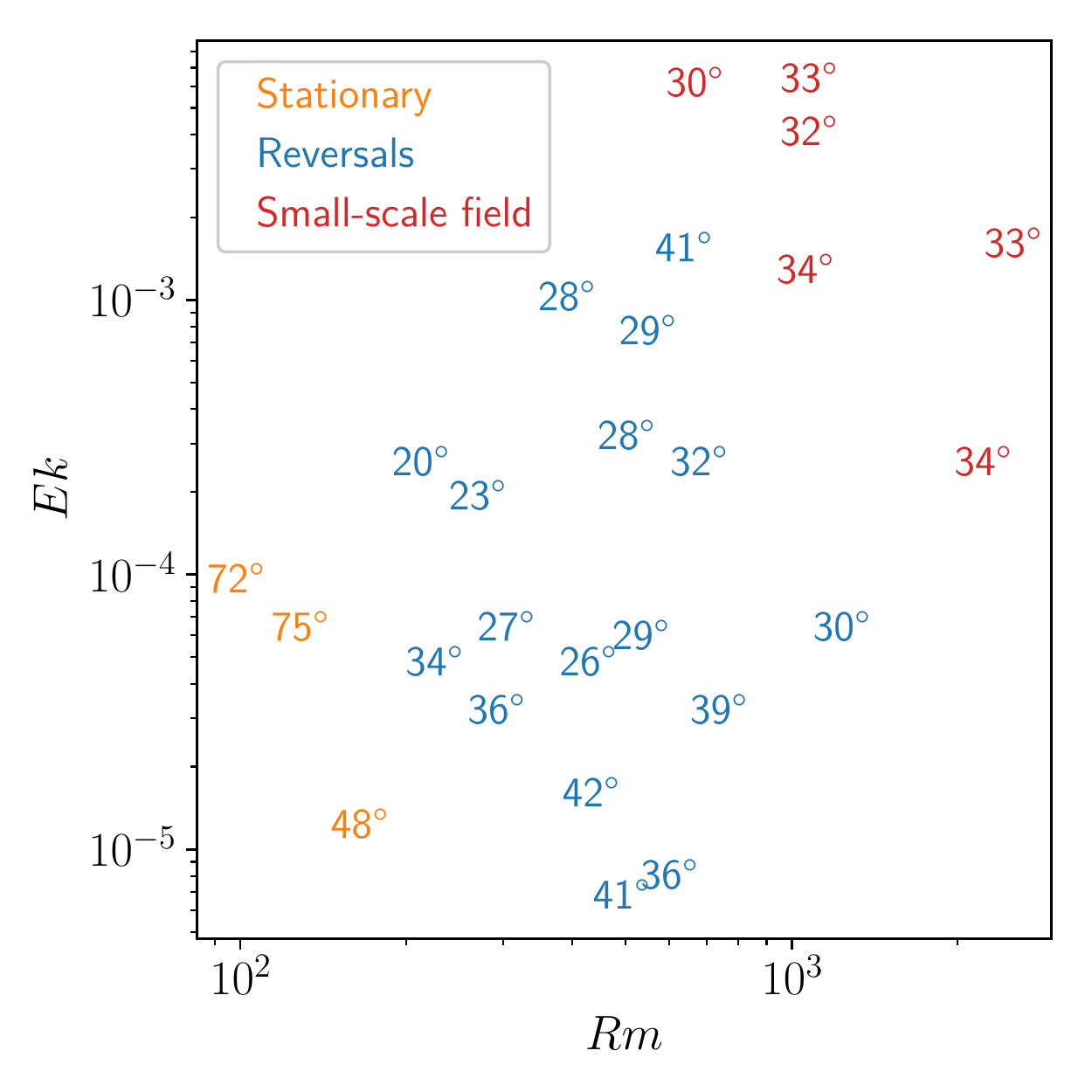}
	\caption{Scatter plot indicating the average absolute latitude of the magnetic dipole moment as a function of the operating Magnetic Reynolds and Ekman numbers. Text color denotes the regime associated to that simulation, using the same labels as in \cref{fig:overview}.}
	\label{fig:tilt}
\end{figure}

Bearing in mind the relative simplicity of our homogeneous MHD dynamo, we study the orientation of $\myvec m$ to further consider similarities and differences between our model and dynamo processes found in planets and laboratory experiments. More precisely, as a raw estimation of the alignment between the rotation axis and the dipole moment, we calculate the time-averaged absolute latitude $\avg{|\alpha|}$, which is shown as a function of Magnetic Reynolds and Ekman numbers in \cref{fig:tilt} (see also table \ref{tab:parameters}).

Considering that a uniformly distributed random dipole vector $\myvec m_R$ has a value $\avg{|\alpha_\text{R}|} \approx \ang{33}$, we conclude that for the small-scale dynamo solutions the dipole moment has no preferential orientation. As $\Ekman$ is decreased to $10^{-4}$, a more equatorial direction for $\myvec m$ (compared with the random distribution) is attained. However, further diminishing $\Ekman$ quickly reverts this trend and the dipole orientation becomes significantly more polar than in the $\myvec m_R$ case. These solutions are more reminiscent of the Earth and celestial dynamos, which are characterized by extremely small values of $\Ekman$ \cite{Roberts2013} and where the rotation axis and the dipole moment are usually aligned \cite{Besse2002}. However, note that laboratory dynamo experiments also display solutions with the magnetic dipole moment in the "equatorial plane" of the experiment, or perpendicular to it, depending on the parameters \cite{Berhanu2010}.

\subsection{Dipole moment variability and 1/f noise}
We now focus our attention on the time variability of the normalized magnetic dipole moment and, in particular, on its component parallel to the rotation axis. Considering that long-time correlations have been reported to be important for this kind of MHD systems \cite{Ponty2004}, the simulations REV12, REV13 and REV14, after being extended up to $2,000$ turnover times, are analyzed now in more detail. These runs were selected because they operate at, approximately, the same value of $\ReynoldsM$ while their Ekman numbers greatly differ. This permits a better study of the effects  the Coriolis force has on the dynamics of the magnetic dipole moment.

\begin{figure}[t]
	\centering
	\includegraphics[width=.45\textwidth,keepaspectratio=true]{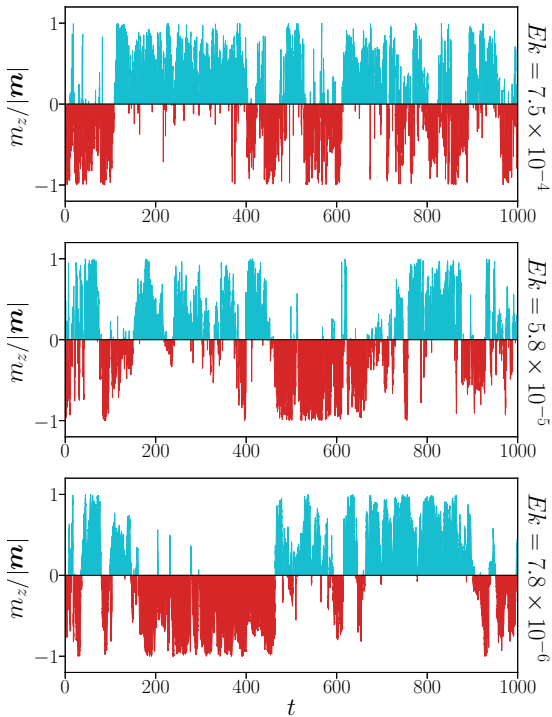}
	\caption{Normalized $z$ component of the magnetic dipole moment as a function of time for runs REV13 (top), REV12 (center) and REV14 (bottom), with decreasing Ekman number. Polarities are marked with cyan when positive and red when negative. Corresponding Ekman numbers are tagged on the right.}
	\label{fig:dipole_time}
\end{figure}

In \cref{fig:dipole_time} we present $m_z/|\myvec{m}|$ as a function of time for the aforementioned simulations. For clarity, only the segment from $t=0$ to $t=1,000$ is displayed. As previously mentioned, the magnetic structure in this region of parameter space displays a dominant dipolar magnetic field, with great variability and several reversals present in each run. The waiting times between polarity inversions $\tau$ seem to spawn a great range of temporal scales, a behavior reminiscent of that observed in paleomagnetic records of the geomagnetic field. Also in \cref{fig:dipole_time} the presence of some excursions can be observed, i.e. abrupt changes in dipole latitude which do not lead to a reversal, a feature also found in the geomagnetic field\cite{Laj2007}. The excursions present in our simulations display varying degrees of latitude variation and spawn a range of timescales comparable to those of the shorter polarity intervals. This fact is most easily appreciated for REV14, although the same conclusion can be drawn for all the reversing simulations.

\begin{figure}[t]
	\centering
	\includegraphics[width=.48\textwidth,keepaspectratio=true]{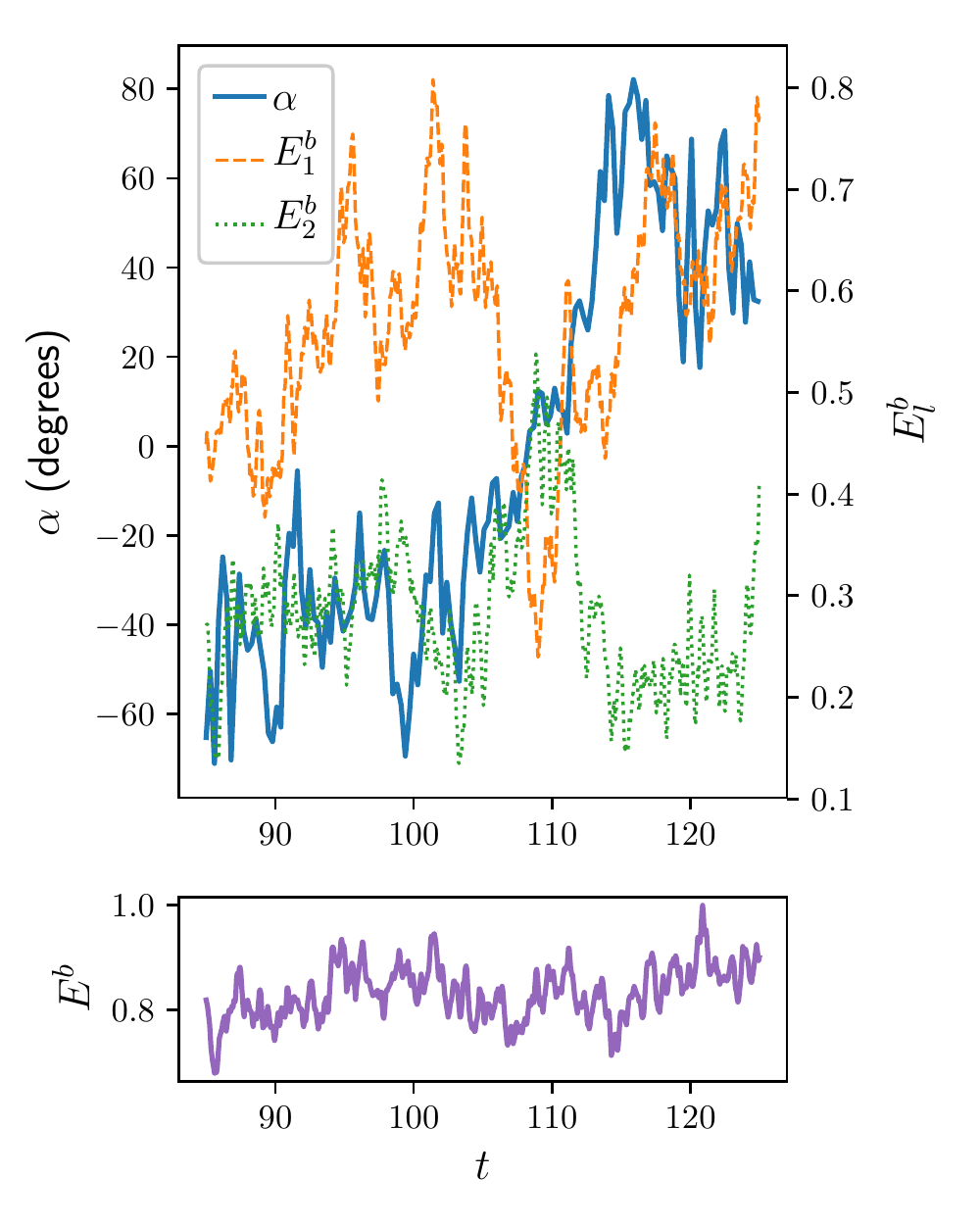}
	\caption{Top: magnetic dipole latitude $\alpha$ (solid blue line), dipolar energy $E^b_1$ (orange dashed line) and quadrupolar energy $E^b_2$ (dotted green line) as a function of time for simulation REV13 during the reversal at ${t\approx108}$. Bottom: total magnetic energy $E^b$ as a function of time for the same run. For convenience, the magnetic energy was normalized to its maximum during the displayed interval.}
	\label{fig:reversal}
\end{figure}

\begin{figure}[t!]
	\centering
	\includegraphics[width=.45\textwidth,keepaspectratio=true]{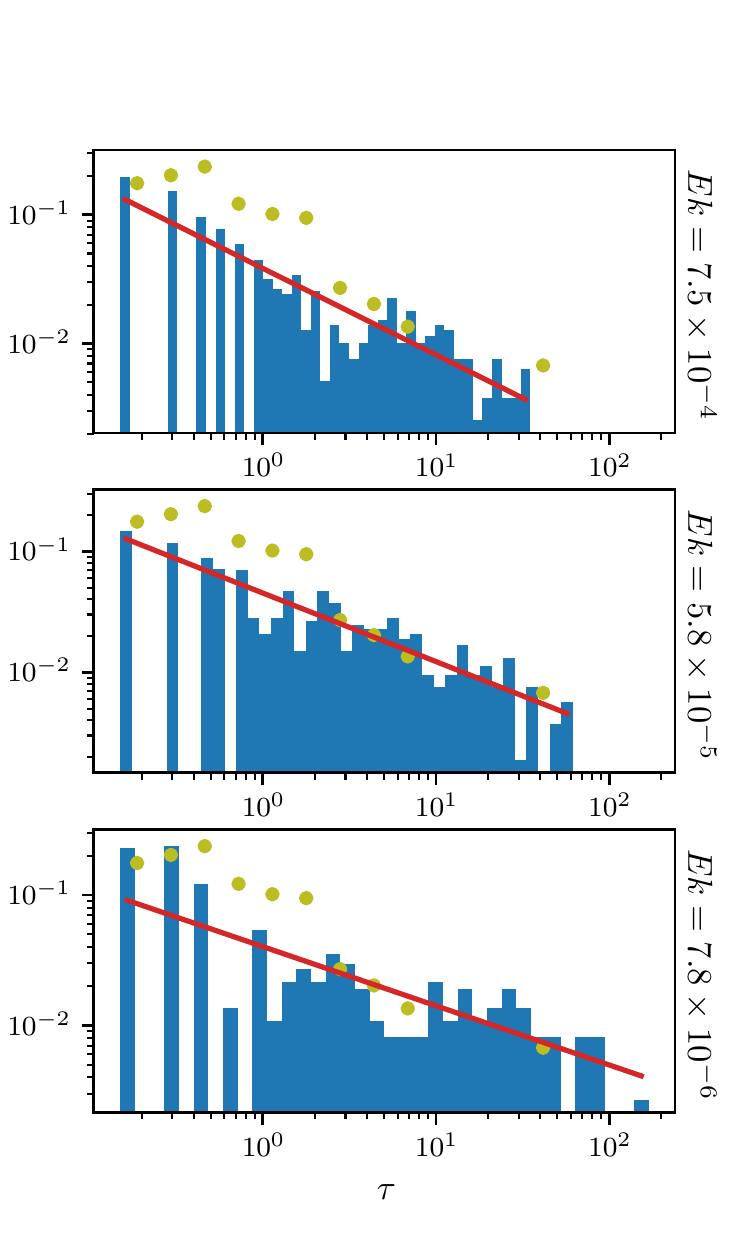}
	\caption{Distribution of the waiting times between reversals for selected simulations (blue bars); the runs are the same as in \cref{fig:dipole_time}. The Ekman number of each run is indicated in the axes. The corresponding mean trend is indicated by the red line. Also, the geomagnetic reversals distribution is shown as a reference by the yellow dots.}
	\label{fig:waiting_t}
\end{figure}

In agreement with previous studies \cite{Amit2010,Leonhardt2007,Olson2011}, we observe an interplay between dipolar and quadrupolar (or higher degree) magnetic energy during the reversal process. This is illustrated in \cref{fig:reversal} where the dipole latitude is displayed in conjunction with the energy contained in the dipolar and quadrupolar magnetic modes before, during and after a typical reversal. It is readily observed that dipolar energy dominates for all times, except for a small interval in the vicinity of the reversal, where the quadrupolar contribution takes over. This behavior is reminiscent of laboratory observations \cite{Monchaux2009,Verhille2010}. Also shown in \cref{fig:reversal} is the magnetic energy normalized to its maximum during the same interval. Although there is some variability, the magnetic energy remains within the same order of magnitude during the whole interval. Moreover, the precise instant of the reversal is not related with a minimum in the magnetic energy.

To better understand the reversals' statistics, the normalized histogram of the waiting times distribution for these runs is calculated and shown in \cref{fig:waiting_t}. Due to the diverse scales involved, a uniform binning in the $\log(\tau)$ domain is employed for the computation. The first thing to note is that simulations with decreasing values of $\Ekman$ seem to be associated with larger maximum values for the waiting times (i.e., with larger probabilities of long times between reversals). However, the most salient point here is that the waiting times follow, approximately, a power-law distribution, implying non-Poissonian statistics and suggesting the presence of long-term memory in the system. This is compatible with the tail of the distribution observed in the data of the geomagnetic field \cite{Cande1995}, also shown as a reference in the same figure. To get comparable dimensional values of $\tau$ between simulations and geomagnetic polarity data, the latter must first be rescaled using proper units. In each panel of \cref{fig:waiting_t}, this is attained by dividing the observational series by its smallest waiting time, and multiplying it in each panel by the smallest waiting time of the corresponding simulation. Note that for the comparison, and as the statistics of $\tau$ are compatible with a self-similar process, we are mostly interested in the power law followed by the probability distribution.

\begin{figure}[t!]
	\centering
	\includegraphics[width=.45\textwidth,keepaspectratio=true]{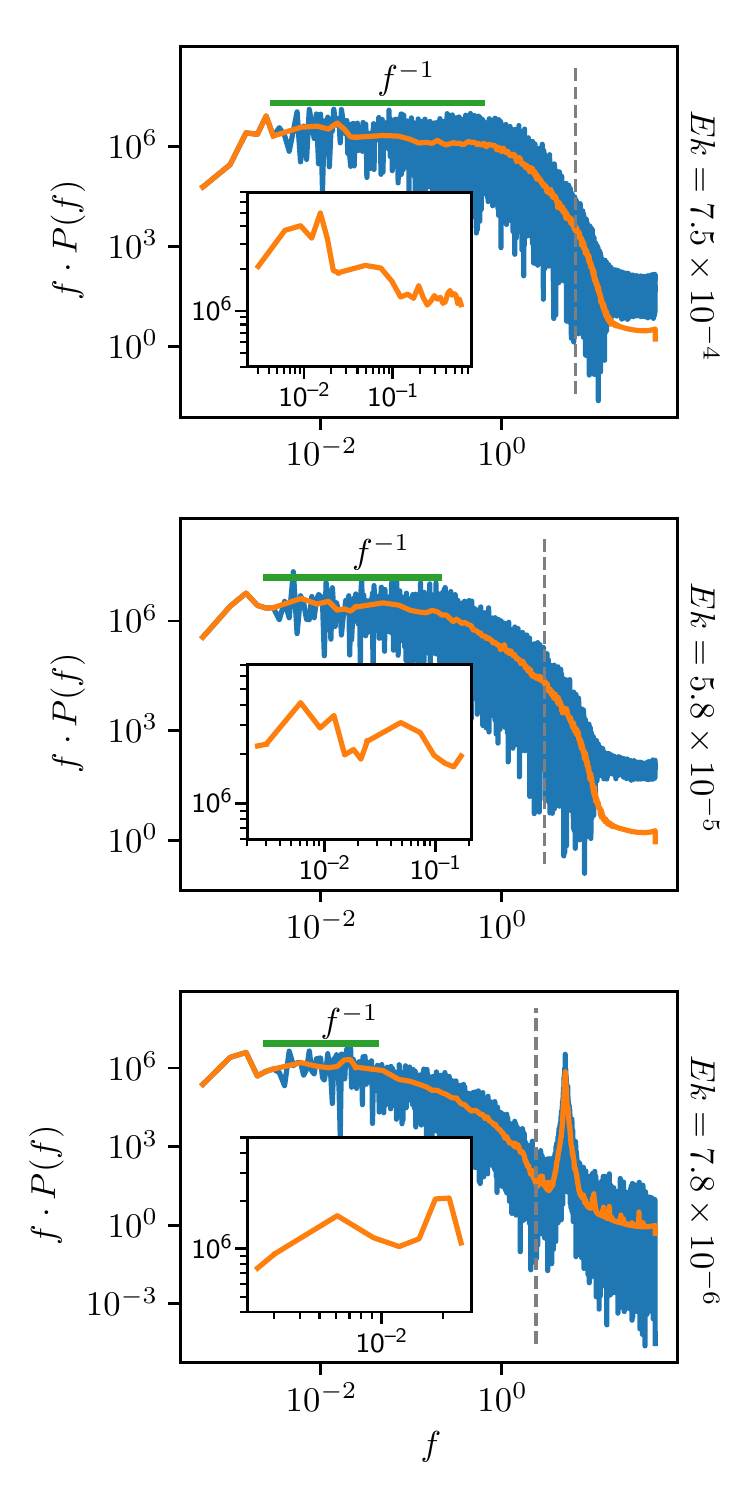}
	\caption{Compensated power spectrum $f P(f)$ for the $z$ component of magnetic dipole moment as a function of the frequency $f$, for the same runs as in \cref{fig:dipole_time}. Both raw (blue) and smoothed (orange) spectra are shown. Regions compatible with $1/f$ behavior are also indicated (green line). Vertical dotted lines mark the energy-containing nonlinear frequency  $f_\text{nl}$. Inset axes display the compensated power spectrum only for the $1/f$ interval. Corresponding Ekman numbers for each run are tagged on the right.}
	\label{fig:dipole_freq}
\end{figure}

Previous research, both numerical and observational, on the geomagnetic dipole moment's power spectrum $P(f)$, has detected a large excess in its magnitude at low frequencies, with an extense region where the spectrum decreases according to a $\sim f^{-1}$ law \cite{Constable2005,Dmitruk2014}, a phenomenon known as $1/f$ noise \cite{Montroll1982,Dutta1981}. Laboratory dynamo experiments also show this behavior \cite{Mininni2014}. In the case MHD dynamos, the development of a $1/f$ spectrum at low frequencies has also been associated with the long term memory in the reversals, and the non-Poissoian statistics of the waiting times \cite{Dmitruk2014}. To study the presence of $1/f$ noise in our simulations, the compensated temporal power spectra $f P(f)$ for simulations REV12, REV13, and REV14 are presented in \cref{fig:dipole_freq}. To compute a smooth spectrum from the raw data (also shown), we apply Welch's method \cite{Welch1967} with two distinct window lengths, in order to adjust both intermediate and high frequencies. All the periodograms can then be rescaled to compensate for the energy lost in the windowing process, resulting in a continuous curve.

For the case of a turbulent MHD flow with an energy containing scale with associated wavenumber $k_0$, and whose interactions are all local in scale, the nonlinear time $t_\text{nl} \approx  1/ [k v(k))]_{k_0}$ is the largest correlation time that can be constructed by triadic interaction in the non-linear term. This corresponds to a frequency $f_\text{nl}=1/t_\text{nl} \approx 4$ in our case (indicated as a reference in \cref{fig:dipole_freq}). Hence, if locality is preponderant, the power spectrum should display a flat response for $f<f_\text{nl}$ (or, in the compensated spectrum $f P(f)$, a line whose slope is proportional to $f$). However, this is not observed in any of our simulations with reversals, for which a great surplus in power for $f < f_\text{nl}$ is found, confirming the presence of long-term memory and indicating the existence of non-local interactions in the system. Moreover, this excess follows (approximately) a $1/f$ law (horizontal line in $f P(f)$, also indicated as a reference in \cref{fig:dipole_freq}). As expected, this $1/f$ signature is absent in stationary dynamos, where Brown noise is observed instead ($P(f)\propto f^{-2}$), but is also present for higher frequencies ($1< f < 10$) in the small scale regime (not shown). Considering that this latter interval is very different from the one found in reversing dynamos in similar regions of parameter space (i.e., it corresponds to frequencies close to the inverse of the eddy turnover time at the forcing scale, and thus it is not a clear signature of a slow process in time), we cannot answer at this point if the mechanisms from which $1/f$ noise emerges is the same in the small scale and reversing regimes.

It is worth analyzing what happens if all possible interactions of triads are considered (even non-local ones). The smallest frequency that arises by means of the triad $(k, q, q)$ in the non-linear term is $f_\text{nl}^q = b(k)/[kv(q)b(q)]|_{k_0}$ (for details, see \cite{Dmitruk2007}). In our case, for the mode $q=3$, we get $f^3_\text{nl} \approx 10^{-1}$, and, if we consider the least local interaction, $q=7$, $f^{7}_\text{nl} \approx 10^{-3}$. Even though this argument can in principle explain how long-term correlations arise in the system, as was previously done in \cite{Dmitruk2007} for ideal systems, the concept of interacting triads seems, however, insufficient to account for the correlations reported in the viscous and resistive case studied here, as $1/f$ intervals appear to be dependent on the operating Ekman number (i.e., on the relative strength of the Coriolis force). Indeed, the data in \cref{fig:dipole_freq} indicates that higher values of $\Ekman$ are associated with the presence of $1/f$ noise at greater frequencies. A related question we cannot answer for the moment, is the extent of the these regions. Sampling points in the lower end of the spectrum are scarce and improving their density is a problem that requires a geometrical increase in the number of timesteps or, equivalently, the required computing power (sampling another decade would require runs ten times longer). Therefore, the extent of the $1/f$ regions as a function of $\Ekman$ remains as an open question.

\section{Conclusions}
With the use of direct numerical simulations we studied the resistive incompressible MHD equations inside a rotating full sphere, including the Coriolis effect, and for a homogeneous flow. Boundary conditions for the magnetic field correspond to an impenetrable and perfect conductor outside the spherical vessel. A very accurate Galerkin spectral method was employed (based on Chandrasekhar-Kendall eigenfunctions), which allowed us to integrate the system equations for thousands of non-linear characteristic times.

We performed thirty runs to do a parametric study varying both Reynolds and Ekman numbers (through variation of the viscosity and of the angular velocity of rotation). We characterized the magnetic field structure of the system according to four different behaviors: no-dynamo, stationary dynamo solutions, solutions with reversal, and small-scale dynamos, regimes with are controlled by the values of the Reynolds and Ekman numbers. The region more reminiscent of solutions found in dynamo experiments and in the geodynamo was the so-called reversal regime, with relatively small Ekman number and high Reynolds number. For that regime we observed hundreds of magnetic dipole reversals, and have been able to perform a statistical analysis of the waiting times between reversals. Our main findings of that analysis were that: (a) non-Poissionian statistics for the waiting times were obtained, with long-time memory, and power law histograms of its distribution, (b) waiting times tend to be longer as the Ekman number is decreased. We also observed a $1/f$ distribution for the frequency ($f$) power spectrum of the magnetic dipole time series, which gives further evidence of an excess power at very low frequencies (longer waiting times), and of a long-term memory process in the system. Consistent with this, the data indicates a shift in the $1/f$ frequency range toward lower frequencies as the Ekman number was decreased, although further confirmation of this trend requires integration of the system very long times and for different values of the Ekman number.

The results build on previous observations of $1/f$ noise and long-term memory in the ideal MHD system \cite{Dmitruk2011}, and in experiments in cylindrical vessels and Taylor-Green numerical simulations \cite{Mininni2014}. The present parametric study in a spherical domain extends these previous works by allowing us to study the statistics of the magnetic dipole moment and of magnetic reversals, and in particular, the identification of trends in their behavior with Ekman and Reynolds numbers. Our results suggest that some long-time statistical properties of dynamo solutions are intrinsic to the MHD system, and can be recovered even with simple configurations or with simplified boundary conditions. Indeed, notwithstanding considerable differences in the operating parameters, our very simplified model of a homogeneous MHD dynamo in a spherical domain yields statistical results compatible with some long-term properties observed in experiments as well as the real geodynamo. Our main limitation is related to the use of a restricted type of boundary conditions, which in turn could explain the relatively low latitude average magnetic dipole obtained (around 40 degrees) in the reversal regime. We plan to improve our model by the use of less-restrictive boundary conditions, while maintaining the desired spectral accuracy, in order to perform long-time simulations.

\section*{Acknowledgements}
The authors acknowledge support from PICT Grant No.~2015-3530 and from PIP Grant No.~11220150100324CO.

\input{main.bbl}
\end{document}